\documentstyle[preprint,aps,epsf]{revtex}

\tightenlines

\begin{document}

\title{
THEORY OF LUMINESCENT EMISSION IN NANOCRYSTAL ZnS:Mn WITH AN EXTRA ELECTRON
}

\author{
{\bf
Nguyen Que Huong and Joseph L. Birman} \\
Physics Department, The City College, CUNY\\
Convent Ave. at 138 St, New York, N.Y. 10031, 
USA} 

\date{\today}
\maketitle

\begin{abstract}

We consider the effect of an extra electron in a doped quantum 
dot $ZnS:Mn^{2+}$. The
Coulomb interaction and exchange interaction between the extra 
electron and the states of the Mn ion will mix the wavefunctions, split the 
impurity energy levels, break the previous selection rules and  change the 
transition probabilities. 
Using this  model of an extra electron in the doped 
quantum dot, we calculate the energy and the wave functions, the luminescence 
efficiency and the transition lifetime and compare with the experiments. 
Our calculation shows that two orders of magnitude of lifetime shortening 
can occur in the transition $^4T_1-^6A_1$, when an extra electron is 
present. 

\end{abstract}
\newpage
{\bf I.Introduction.}
  
   In contrast to undoped materials, the impurity states in a doped 
nanocrystal play an 
important role in the electronic structure, transition probabilities and the optical 
properties. 
In recent years, attempts to understand more about these zero-dimensional nanocrystal 
 effects have been made in several 
labs by doping an impurity in a nanocrystal, searching for novel 
materials and new properties, and among them Mn-doped ZnS nanoparticles have been intensively studied 
[1-15]. Among many bulk wide band gap compounds, manganese is well known 
as an activator for 
photoluminescence 
(PL) and electroluminescence (EL) and the $Mn^{2+}$ ion d-electrons states act as 
efficient luminescent centers while doped into a semiconductors. 

In 1994 Bhargava and Gallagher[1, 2] reported the first realization of a ZnS 
semiconductor 
nanocrystal doped with Mn isoelectronic impurities and claimed that Mn-doped ZnS 
nanocrystal can yield both high luminescence efficiency and 
significant lifetime shortening.
The yellow emission characterized for $Mn^{2+}$ in bulk ZnS [16-18],
which is
associated with 
the transition $^4T_1 - ^6A_1$, was reported to be observed in photoluminescence (PL) spectra 
for the $Mn^{2+}$  in nanocrystal ZnS. 
In nanocrystals, however, the PL peak for the yellow emission is reported slightly shifted 
toward a  lower energy (in bulk ZnS:Mn it peaks arount 2.12 ev, in nanocrystal 
ZnS:Mn it peaks at 2.10 eV).
Also, the reported linewidth of the yellow emision in the PL spectrum for a nanocrystal is larger 
than for the bulk.
Most strikingly, the luminescence lifetime of the $Mn^{2+}$ $^4T_1-^6A_1$ 
transition was reported to 
decrease by 5 orders of magnitude, from 1.8 ms in bulk to 3.7 ns and 20.5 ns in nanocrystals
while maintaining the high ($18\%$) quantum efficiency.

In ref. [6] the authors suggested that the increase in quantum efficiency as well as the 
lifetime 
shortening is the result of strong hybridization of s-p electrons of the ZnS host and d-
electrons of the Mn impurity due to confinement, and also of the modification of the crystal 
field near the surface of the nanocrystals.  
Stimulated by this dramatic result, many other laboratories are 
trying to synthesize the Mn-doped ZnS nanocrystals and considerable attention has 
been paid to optical properties of these kind of materials[7-13]. 
Yu et al.[8] reported obtaining the yellow emission peak of $^4T_1-^6A_1$,
the slight shift of the peak toward the high energy and the increase in luminescence efficiency.
Sooklai et al. [9] reported the shortening of the decay time to ns for nanocrystals $ 
ZnS:Mn^{2+}$ while Ito et al. [10] obtained the life time shortening to $\mu s$
in $ZnTe:Mn^{2+}$ quantum well. 
Very recently, D. Norris et al.[13] reported obtaining high quality 
ZnSe colloidal nanocrystals doped with single $Mn^{2+}$ impurities.
Also, several different 
groups have proposed models to try to explain the processes occuring with 
the impurity centers inside the confined nanocrystals [6,14-15].
But while it is now well established
that the confinement effect strongly modifies the electronic structure of 
nanocrystals,  the effects of the confinement on the 
energy structure and transitions of the Mn impurity in the nanocrystals, still
are controversial. Also, there might exist different mechanisms to change the optical
processes for doping centers in nanocrystals.

Here we investigate a new effect and mechanism to study Mn doping in a nanocrystal.
Our model is to assume there is an extra electron in a doped dot.
The electron may originate in the dot itself, or might be injected into the dot. 
The extra electron will strongly
couple to the electron involved in the transitions, split
the energy levels and mix the wavefunctions, and will thus 
break the former selection rules. 
Because the dot is small, the boundary conditions enhance the coupling, 
so by controlling the presence or absence of the extra
electron in the dot, one can control the optical transitions
in the dot.
This can provide the explanation for the observed shortening of lifetime and
enhancement of the quantum efficiency.

The transition of our interest here is $^4T_1-^6A_1 $ of the $Mn^{2+}$ ion
in the crystal field of the ZnS nanocrystal. The luminescent transition
from the lowest excited level $^4T_1$ (spin $\normalsize\frac{3}{2}$) to
the ground state (spin $\normalsize\frac{5}{2}$) is spin-forbidden.
But, the weak spin-orbital interaction makes the transition slightly allowed
[16-18].

Actually, for the Mn-center we have a configuration of $d^5$ electrons. If we consider 
the ways in which transitions can take place between the energy levels
of a $d^n$ center, we find that the electric dipole transitions are
forbidden due to parity ( all the d-states are of even parity). 
But for the crystal field whose symmetry group 
does not contain the inversion symmetry, which is our case (local symmetry is $C_{3v}$ in
the wurtzite symmetry of $ZnS$), the electric 
dipole transition becomes possible due to to the mixing of odd-parity states 
into the $d^n$ states, which means the wavefunctions of the Mn-center states now
contains both the parities [19-20].Meanwhile, a small fraction of the 4p 
atomic orbitals is 
likely mixed into the 3d-orbital and the typical states are written in the form:
\begin{equation}
\Psi^i(3d^n) = \Psi^i_0(3d^n) + \beta\chi^i(4p)
\end{equation}

\noindent
where $\Psi^i_0(3d^n)$ is the i-state even parity wavefunction of the $d^5$ electron and $\chi^i(4p)
$ is the odd parity wavefunction of the $4p$ electron. $\beta$ is a small coefficient.
Then the electric dipole transitions between the i-state and j-state, for example, of the doping center 
can arise due to non-zero matrix elements between $\Psi^i_0(3d^n)$ and $\epsilon\chi^j(4p)$ and between 
$\epsilon\chi^i(4p)$and $\Psi^j_0(3d^n)$ even though the matrix elements between $\Psi^i_0(3d^n)$ and
$ \Psi^j_0(3d^n)$ are zeros. 

Due to this unsymmetrical environment the electric dipole transition between the 
ground state $^6A_1$ and the first excited 
state $^4T_1$ is orbitally possible.But due to spin difference the transition is spin-forbidden.
This rule is slightly relaxed due to the spin-orbit interaction. 
But because the spin-orbit interaction is very 
weak, the oscillator strength is very small , which leads to the luminescence life time of ms order 
of 
magnitude. In order for the transition to be allowed, there must be some mixing that breaks the spin 
selection rules.

If an extra electron is in the dot, the electron will couple with the 
electrons of $Mn^{2+}$ in each state, namely the 
electron will couple with
the states $^4T_1$ and $^6A_1$. The coupling will mix different states and will produce the
states with the same spin, between which the electric dipole will be allowed.

In this work we consider as the coupling hamiltonian 
the Coulomb and Exchange interactions. The pertubed wavefunctions of the $d^5$ electrons
in $^6A_1$ and $^4T_1$ states will be derived in the next two sections.
For the  extra electron which is confined in the dot and localized at 
some lattice point, we will use the Wannier exciton function:
\begin{equation}
\Psi^{Dot}_{1S}(r) = \sum_{a_i}\varphi^{Dot}_{1S}(r)w_{a_i}(r-a_i)
\end{equation}

\noindent
here $w_{a_i}(r-a_i)$ is the Wannier function of the exciton localized at the lattice 
point $a_i$, where $a_i$ is the distance from the origin to the lattice point i inside the dot. 
$\varphi^{Dot}_{1S}(r)$ is the envelope function for a confined electron in the sphere:
\begin{equation}
\varphi^{Dot}_{(1s)} = R_{1s}^{Dot} Y^0_0|\frac{1}{2},\sigma>
\end{equation}

\noindent 
where $Y^0_0$ is a spherical harmonics, $|\frac{1}{2},\sigma>$ is the spin function
for the electron with spin $s=\normalsize\frac{1}{2}$ and $\sigma$ = + or - for
spin up and spin down states. $R_{1s}^{Dot}$ is the envelope function for the
$1S$ electron confined in a sphere of radius R (2.4)
\begin{equation}
R_{nl}^{Dot}(r) = \sqrt{\frac{2}{R^3}}\frac{j_l(\chi_{nl}\frac{r}{R})}
{j_{l+1}(\chi_{nl})}
\end{equation}

\noindent
where $j_{nl}$ is the spherical Bessel function,$\chi_{nl}$ is the
location of the zero of the spherical Bessel function.

The interaction Hamiltonian between the extra electron and the $d^5$
electrons is given in section III. In section II we will derive the wavefunctions for the 
$d^5$ electrons in $ ^4T_1 $ and $^6A_1$ states using the strong limit field approximation.
\vspace{0.5cm}

{\bf II. $3d^5$ Electron in the Crystal Field -Crystal Field.  Manganese Energy Levels and
Wavefunctions}

  Impurity centers are formed by foreign ions substituting for host ions or placed interstitially in 
the lattice. 
When the activator ions such as Mn are placed in the crystal field of the 
semiconductor lattice, the crystal field would affect the wavefunctions
and energy structure of the impurity to form " crystal field states" of the 
impurity ions.
For the transition metal ions with the outer $3d^n$ electron configuration, the crystal field 
energy and the interelectron Coulomb interaction are comparable[18,19], 
so these $d^n$ electrons 
can be treated either in the intermediate crystal field or in the strong crystal field limit, and they are 
more often
treated in the strong field approximation.
Then the fivefold
degenerate 3d level will split principally into two levels, the two-fold degenerate level with the 
additional energy +6Dq, and the three-fold degenerate one with the additional energy of 
-4Dq.
Here Dq is a single parameter which characterizes the strength of an 
octahedral crystal field [19,20].
In Fig. 1 the splitting diagram for a 3d electron in the octahedral crystal field 
is shown[19]. 
In this section we will derive the  wavefunctions of the
$d^5$ electrons of the Mn impurity in the crystal field.
We note that it is customary in this system to treat the local site symmetry as octahedral for the major 
$t_2-e$ splitting and then add a small axial field to reduce the site symmetry
to the $C_{3V}$ of wirtzite structure.

The eigenstates of the two-fold degenerate level ( the e-orbitals) are written in the 
following form (see eqn (1) above for prototype):
\begin{eqnarray}
\phi_{eu} & = & |3d0> = R_{3d}(r)(\frac{5}{4\pi})^{1/2}
(\frac{3z^2-r^2}{2r^2}), \nonumber \\ 
\phi_{ev} & = & (\frac{1}{2})^{1/2}(|3d2> + |3d-2>) = 
R_{3d}(r)(\frac{5}{4\pi})^{1/2}3^{1/2}(\frac{x^2-y^2}{2r^2})
\end{eqnarray}

\noindent
And the eigenstates for the three-fold degenerate level (the $t_2$-orbital) 
are written as:
\begin{eqnarray}
\phi_{t_2\xi} & = & (\frac{i}{2})^{1/2}(|3d1> + |3d-1>) =
R_{3d}(r)(\frac{5}{4\pi})^{1/2}3^{1/2}(\frac{yz}{r^2}), \nonumber \\
\phi_{t_2\eta} & = & -(\frac{1}{2})^{1/2}(|3d1> - |3d-1>) =
R_{3d}(r)(\frac{5}{4\pi})^{1/2}3^{1/2}(\frac{xz}{r^2}), \nonumber \\
\phi_{t_2\zeta} & = & -(\frac{i}{2})^{1/2}(|3d2> - |3d-2>) =
R_{3d}(r)(\frac{5}{4\pi})^{1/2}3^{1/2}(\frac{xy}{2r^2})
\end{eqnarray}

\noindent
For simplicity, from now on we will denote these double degenerate and triple degenerate
eigenstates $ u,v$ and $\xi$, $\eta, \zeta$, respectively.

For a multi-3d-electron system, the orbital Hamiltonian in crystal field will be
the following:
\begin{equation}
H = \sum_{i}{H^0_0(r_i)} + \sum_{i>j}{H'(r_i, r_j)}
\end{equation}

\noindent
where $H^0_0(r_i)$ is the Hamiltonian  of a single electron interacting
with crystal field , $H'(r_i, r_j)$ is the electron-
electron interaction Hamiltonian between the d-electrons.
For the strong crystal field limit, the Hamiltonian  $H^0_0(r_i)$ will
be solved first to obtain the wavefunctions of a single 3d electron as in the last 
chapter, then these states will be interacting by the Hamiltonian $H'(r_i, r_j)$
The wavefunctions of 3d-electron orbitals are $ u,v$, which belong to
the E irreducible representation, and $\xi,\eta,\zeta$ which belong to
the $T_2$ reducible representation.
So in the strong crystal field case, the wavefunctions of
multi-3d-electron Hamiltonian will be the products of these one-electron
orbitals.

In an octahedral crystal field, the Hamiltonian (5) is invariant under all the rotation
operators of the group $O_h$, where for the first term the rotations are 
applied independently for each $r_i$ and for the  second term the 
rotations are applied simultaneously for all $r_i$.
Then the wavefunctions will transform according to irreducible representation
of the $O_h$ group.

 Begining the reduction process for the two 3d electrons configuration, we will
have the products $E\times E, T_2\times T_2$ and $T_2\times T_2$ [19,20].
The $E\times E$
product functions will belong to $A_1, A_2$ and E irreducible representation.
We will denote as $|e^2, A_1>$ the function of the product $E\times E$ which belongs to the 
$A_1$ representation.
The $|e^2, A_1>$ function is symmetric under interchange of $r_1, r_2$. Then because
of the Pauli principle, to obtain the antisymmetric total function its spin function
must be the antisymmetric spin function with the total spin S=0
\begin{equation}
|S=0, M_S=0> = \frac{1}{\sqrt{2}}\left[\alpha(\sigma_1)\beta(\sigma_2) -
\beta(\sigma_1)\alpha(\sigma_2)\right]
\end{equation}

\noindent
where $\alpha(\sigma)$ and $\beta(\sigma)$ are the up ($|\uparrow>$)
and down ($|\downarrow>$) spin functions. 
If one use the wavefunction $[u^+u^-]$ to denote 
the normalized two by two Slater determinant of 
the products of the orbital and
the spin functions $u(r_i)\alpha(\sigma_j)$ where $i,j = 1,2$ then the final full 
wavefunction $|e^2, A_1>$ is written in the short form
\begin{equation}
|e^2,A_1, M_S=0> =  \frac{1}{\sqrt{2}}\left([u^+u^-] + [v^+v^-]\right)
\end{equation}

\noindent
Similarly for the wavefunctions $|e^2,A_2>$ of the $E\times E$ orbitals which belongs
to $A_2$ representation  we have
the antisymmetric orbital wavefunction, then the spin functions must be three
symmetric spin functions of the total spin S=1. Denoting the spin index $2S+1$
on the left of the representation $A_2$ we write the total wavefunctions 
of $|e^2, A_2>$ in the form:
\begin{eqnarray}
|e^2, ^3A_2, M_S=1> & = & |u^+v^+| \nonumber \\
|e^2, ^3A_2, M_S=0> & = & \left([u^+v^-] + [u^-v^+]\right) \nonumber \\
|e^2, ^3A_2, M_S=-1> & = & [u^-v^-]
\end{eqnarray}

\noindent
and for the $|e^2,E>$ wavefunctions one gets:
\begin{eqnarray}
|e^2,^1Eu, M_S=0> & = & \frac{1}{\sqrt{2}}\left([u^+u^-] - [v^+v^-]\right)\nonumber \\
|e^2,^1Ev, M_S=0> & = & \frac{1}{\sqrt{2}}\left([u^+v^-] - [u^-v^+]\right)
\end{eqnarray}

\noindent
Continuing in this way one can get all the possible wavefunctions
and energies for $ t^n,
e^n$ and $t^me^k$ configuration of $3d^n$ in the octahedral crystal
field [20].
The ground state of the $3d^5$ configuration is the sextet $^6A_1$ with the total 
spin $\normalsize\frac{5}{2}$.
The first excited level is the triplet states $^4T_1$, which is the excited state 
closest to the ground state. 
We are interested in the optical transition between the first 
excited state and the ground state, this is the transition $^4T_1 -^6A_1$. 
The wavefunction of $^6A_1$ and $^4T_1$ states can be derived, using the table of 
coupling coefficients from [20,21,23]. 

The ground state - the sextet $^6A_1$ is derived from $t_2^3 \times e^2$. From the 
table [21] the $^6A_1$ state is:
\begin{eqnarray}
|^6A_1> & = & \left|t_2^3 (^4A_2) e^2(^3A_2), ^6A_1a_1 \right\rangle \nonumber \\
&  = & \left|t_2^3 (^4A_2)\right\rangle \left|e^2 (^3A_2)\right\rangle
\end{eqnarray}

\noindent
The functions of $t_2^3(^4A_2)$ and $E^2(^3A_2)$ are [21]:
\begin{eqnarray}
|t_2^3 (^4A_2) \frac{3}{2}a_2> & = & [-\xi^+\eta^+\zeta^+] \nonumber \\
|e^2 (^3A_2) 1a_2> & = & [u^+v^+]
\end{eqnarray}

\noindent
Then we have for the wavefunctions of the ground state $^6A_1$:
\begin{equation}
|^6A_1a_1> = [-\xi^+\eta^+\zeta^+u^+v^+]
\end{equation}

\noindent 
The triplet $^4T_1$ wavefunctions of $3d^5$ are derived from $t_2^4(^3T_1) \times e^1(^2E)$.
\begin{eqnarray}
|^4T_1(3d^5) \frac{3}{2}x> & = & |t_2^4(^3T_1)e^1(^2E), ^4T_1 \frac{3}{2}x>\nonumber \\
|^4T_1(3d^5) \frac{3}{2}y> & = & |t_2^4(^3T_1)e^1(^2E), ^4T_1 \frac{3}{2}y>\nonumber \\
|^4T_1(3d^5) \frac{3}{2}z> & = & |t_2^4(^3T_1)e^1(^2E), ^4T_1 \frac{3}{2}z>
\end{eqnarray}

\noindent
and we obtained the wavefunctions of the triplet $^4T_1$ of the 
$d^5$ configuration in the octahedral crystal field:
\begin{eqnarray}
|^4T_1(3d^5)\frac{3}{2}x> & = & -\frac{1}{2}[\xi^+\xi^-\eta^+\zeta^+u^+] -
\frac{\sqrt{3}}{2}[\xi^+\xi^-\eta^+\zeta^+v^+] \nonumber \\
|^4T_1(3d^5)\frac{3}{2}y> & = & \frac{1}{2}[\xi^+\eta^+\eta^-\zeta^+u^+] +
\frac{\sqrt{3}}{2}[\xi^+\eta^+\eta^-\zeta^+v^+] \nonumber \\
|^4T_1(3d^5)\frac{3}{2}z> & = & [\xi^+\eta^+\zeta^+\zeta^-u^+] 
\end{eqnarray}

\noindent
We recall here that the wavefunctions in (14) and (16) are
five by five Slater determinants which we have written in the
short notation. 

As discussed before, in the actual relevant $C_{3V}$ crystal field ZnS:Mn with no 
inversion symmetry, the $4p$-states 
with odd parity can be mixed into the $3d^5$ state with even parity. Consider the 
configuration of the $3d^5$ electrons of the Mn- center with the $^6A_1$ being the ground 
state and $^4T_1$ as the first excited state. We can assume that because the $^6A_1$ level 
is the ground state, it can be considered to be far away from other states, including the 4p state. 
So we can consider the $^6A_1$ as a $3d^5$ -pure state. Since the $^4T_1$ is the excited state, it will
be closer in energy to other 4p state and the probability that it will be mixed with the odd-parity
states will be much higher. The p electron has the $t_1$ wavefunction. From the configuration 
of $3d^5$ electrons, we see that the $^4T_1$ state arose from $t_2^4(^3T_1) \times e^1(^2E)$. 
Because of symmetry, it is reasonable to assume that the state 
$|3d^5\{^4T_1 (t_2^4 \times e^1)\}>$ is more likely to mix with the configuration 
$|3d^44p\{^4T_1(t_2^4 \times t^1)\}>$. Or we can say that the configuration with four d-electrons in 
the function $t_2$ and one electron in function e is more likely to mix with the configuration with only a little 
energy difference, which also has 4 d-electrons in the same functions $t_2$ and only the fifth electron 
being the $t_1$ function instead of the e-function as in other configuration.   

Now consider the configuration $|3d^44p\{t_2^4 \times t^1)\}>$. The $t_2^4$ configuration is a combination of 
$^1A_1 + 
^1E +^3T_1 +^1T_2$. The product of all these four functions with the function $^2T_1$ of the p-electron can give 
the  triplet $T_1$, but among those states only the product of $^3T_1$ with $^2T_1$ can give the exact spin of 
the state $^4T_1$. 
Then we have the only odd state which can mix with our $|3d^5\{^4T_1 [t_2^4(^3T_1) \times e^1(^2E)]\}>$ state is 
the state $|3d^44p\{^4T_1(t_2^4(^3T_1) \times t^1(^2T_1))\}>$.

Recalling the eigenstates of the p-electron:
\begin{eqnarray}
|p_z> & = & |2P0> = -iR_{20}|Y_{10}>, \nonumber\\
|p_x> & = & \frac{i}{\sqrt{2}}(|2P1>-|2P-1>) = \frac{i}{\sqrt{2}}R_{20}(|Y_{11}>-|Y_{11}>), \nonumber\\
|p_x> & = & \frac{1}{\sqrt{2}}(|2P1>+|2P-1>) = \frac{i}{\sqrt{2}}R_{20}(|Y_{11}>+|Y_{11}>)\nonumber \\
\end{eqnarray}

\noindent
we obtain the triplet $^4T_1$ of the $d^4p$ configuration in the crystal field:
\begin{eqnarray}
|^4T_1(3d^44p)x> & = & \frac{1}{\sqrt{2}}[\xi^+\eta^2\zeta^+p_z^+] -
\frac{1}{\sqrt{2}}[\xi^+\eta^+\zeta^2p_y^+] \nonumber \\
|^4T_1(3d^44p)y> & = & \frac{1}{\sqrt{2}}[\xi^2\eta^+\zeta^+p_z^+] -
\frac{1}{\sqrt{2}}[\xi^+\eta^+\zeta^2p_x^+] \nonumber \\
|^4T_1(3d^44p)z> & = & -\frac{1}{\sqrt{2}}[\xi^2\eta^+\zeta^+p_y^+] -
\frac{1}{\sqrt{2}}[\xi^+\eta^2\zeta^+p_x^+]
\end{eqnarray}

\noindent
And finally we have  the wavefunctions  of the triplet $^4T_1$ of the Mn-center:
\begin{eqnarray}
|^4T_1x> & = & -\frac{1}{2}[\xi^+\xi^-\eta^+\zeta^+u^+] -
\frac{\sqrt{3}}{2}[\xi^+\xi^-\eta^+\zeta^+v^+] + \frac{\beta}{\sqrt{2}}[\xi^+\eta^2\zeta^+p_z^+] -
\frac{\beta}{\sqrt{2}}[\xi^+\eta^+\zeta^2p_y^+] \nonumber \\
|^4T_1y> & = & \frac{1}{2}[\xi^+\eta^+\eta^-\zeta^+u^+] +
\frac{\sqrt{3}}{2}[\xi^+\eta^+\eta^-\zeta^+v^+] + \frac{\beta}{\sqrt{2}}[\xi^2\eta^+\zeta^+p_z^+] -
\frac{\beta}{\sqrt{2}}[\xi^+\eta^+\zeta^2p_x^+] \nonumber \\
|^4T_1(3d^5)\frac{3}{2}z> & = & [\xi^+\eta^+\zeta^+\zeta^-u^+]  -\frac{\beta}{\sqrt{2}}[\xi^2\eta^+\zeta^+p_y^+] 
-\frac{\beta}{\sqrt{2}}[\xi^+\eta^2\zeta^+p_x^+]
\end{eqnarray}

{\bf III. The Coulomb and Exchange Interaction}

The Coulomb and Exchange interaction will couple the extra electron and the impurity  
electrons and then result in different spin configurations, so we expect this 
will make the spin-forbidden transition become allowable. 
We call the Coulomb integral[24,25]
\begin{equation}
K(a_1b_1, a_2b_2) = <a_1,b_1|V_{12}|a_2b_2>
\end{equation}

\noindent
and the exchange integral
\begin{equation}
J(a_1b_1, a_2b_2) = <a_1,b_1|V_{12}|b_2a_2>
\end{equation}

\noindent
The exchange interaction 
between the extra electron and the impurity configuration has the following form:
\begin{equation}
H_{ex} = -JS_{Mn} S_e
\end{equation}

\noindent
where $S_{Mn}$ is the total spin of the $d^5$ configuration in the states
of interest, $S_e$ is spin of the injected electron.
J is the exchange matrix element, which has the form:
\begin{equation}
J_{ii} =\sum_{k,\lambda}{ <\overline{\phi}_k\overline{\phi}_{\lambda}| V |
\phi_{\lambda}\phi_k> }
\end{equation}

\noindent
where $\phi_k$  is the electron wavefunction, V is the two-electron
interaction operator:
\begin{equation}
V (r_1,r_2) = \frac{q_1q_2}{r_{12}}
\end{equation}

\noindent
Because both the wavefunctions of the extra electron and of
the $d^5$ electrons in the $^6A_1$ and $^4T_1$ state 
are expressed in terms of spherical harmonics,
it will be natural that for the two-electron interaction
operator V we will use the expression in terms of the spherical harmonics.
V can be expanded in the form[20]:
\begin{equation}
V (r_1, r_2)= \frac{q_1q_2}{r_{>}}\sum_{k=0}^{\infty}{\frac{4\pi}{2k+1}\frac{r^k_{<}}
{r^k_{>}}}\sum_m{Y_{km}(1)\overline{Y}_{km}}(2)
\end{equation}

\noindent
If we put the electron functions into equations (20) and (21), the matrix elements
will consist of two parts - the radial part and the angular part:
\begin{equation}
<\phi_i\phi_j| V | \phi_m\phi_t> = \sum_{k=0}^{\infty}{\rho^k(n^il^i,n^jl^j,
n^ml^m,n^tl^t) \times A_k\sigma(s^is^t)\sigma(s^js^m)}
\end{equation}

\noindent
where $\rho^k(n^il^i,n^jl^j,n^ml^m,n^tm^t)$ is the radial part
\begin{equation}
\rho^k(n^il^i,n^jl^j,n^ml^m,n^tm^t) = <R_{n^il^i}R_{n^jl^j}|\frac{e^2r^{k}_{<}}
{r^{k+1}_{>}}|R_{n^ml^m}R_{n^tl^t}>
\end{equation}

\noindent
and $A_k$ is the angular part
\begin{equation}
A_k = \frac{4\pi}{2k+1}\sum_{p=-k}^{+k}
{\langle\overline{Y}_{l^ip^i}Y_{kp}Y_{l^tp^t}\rangle
\langle\overline{Y}_{l^jp^j}\overline{Y}_{kp}Y_{l^mp^m}\rangle}
\end{equation}

\noindent
where $\langle{\overline{Y}_{l^ip^i}Y_{kp}Y_{l^tp^t}}\rangle$ is a
Clebsch-Gordan coefficient, which is different from zero only when the following
conditions are satisfied:
\begin{eqnarray}
p^i & = & p+p^t \nonumber\\
k + l^i + l^t & = & even   \nonumber\\
l^i - l^t & \leq & l^i+l^t
\end{eqnarray}

\noindent
These conditions will reduce the terms in the sum and leave only several
matrix elements different from zero.

Now with all the wavefunctions and the interaction form we are ready 
to calculate the Coulomb and Exchange interaction between
the extra electron and the electrons in Mn-center.
\vspace{0.5cm}

{\bf IV. Exchange Interaction of $Mn^{2+}$ Ion with the Extra
Electron in the Nanocrystal}

Now we suppose that the extra electron inside the nanocrystal is close
enough to the $Mn^{2+}$ ion at the center so that the electron can couple
via exchange interaction with the $d^5$ electrons in both the ground
state $^6A_1$ and the excited state $^4T_1$. In this section we will consider the effect of this
exchange interaction on the impurity electrons in states $^6A_1$ and $^4T_1$. 

The matrix element of the exchange interaction of the extra electron
and the impurity electrons in the $^6A_1$ state is:
\begin{eqnarray}
& & \left\langle^6A_1(\vec{r}) \Psi^{Dot}_{1s}(\vec{r'})| V(r-r')|\Psi^{Dot}_{1s}
(\vec{r})^6A_1(\vec{r'})\right\rangle \nonumber\\
\mbox{} \nonumber \\
& & = \left\langle\overline{[-\xi^+\eta^+\zeta^+u^+v^+](r)}
\overline{\varphi^{Dot}_{1s}(r')Y^0_0}\left|
V(r-r')\right|\varphi^{Dot}_{1s}(r)Y^0_0 [-\xi^+\eta^+\zeta^+u^+v^+](r)\right\rangle
\nonumber\\ 
\end{eqnarray}

\noindent
We should pay attention here to the point that the wavefunction of the 
$^6A_1$
state of $d^5$ electrons is a Slater determinant with five single d-electron
wavefunctions.
So the coordinate
$\vec{r}$ of the $d^5$ electron wavefunction is actually five
different coordinates $(r_1, r_2,r_3, r_4, r_5)$ of these five single
d-electron wavefunctions. In fact, the extra electron 
has an exchange interaction with each of the five d-electrons.
Because these five coordinates are independent, the 
integrals will be taken separately.

In the calculation of the determinantal matrix elements of (30), we have
to deal with the single electron matrix element of the type, for example:
\begin{eqnarray}
& &\left\langle\xi(r)\varphi(r')|V(r-r')|\varphi(r)\xi(r')\right\rangle \nonumber \\
\mbox{} \nonumber \\
& & =  \frac{4\pi q_1q_2}{j_1(\pi)^2}\frac{1}{R^3}\sum_{k=0}^{\infty}\frac{1}{2k+1}
\left\langle\overline{R_{3d}(r)\sum_{a_i}{w_{1S}(r-a_i)}}\left(\overline{Y^1_2} + \overline{Y^{-1}_2}\right)
\overline{j_0}\left(\pi \frac{r'}{R}\right)\overline{Y_0^0} \right| 
\times \nonumber\\
\mbox{} \nonumber \\
& & \frac{r^k}{r'^{k+1}}\sum_{m=-k}^k{Y_k^m(1)Y_k^m(2)} \left|
j_0\left(\pi\frac{r}{R}\right)Y_0^0R_{3d}(r')\sum_{a'_i}{w_{1S}(r'-a'_i)}\left(Y_2^1 + Y_2^{-1}\right)
\right\rangle
\end{eqnarray}

\noindent
or
\begin{eqnarray}
& &\left \langle\xi(r)\varphi(r')|V(r-r')|\varphi(r)\xi(r')\right\rangle \nonumber \\
\mbox{} \nonumber \\
& &  = \frac{4\pi q_1q_2}{j_1(\pi)^2}\frac{1}{R^3}\sum_{k=0}^{\infty}\frac{1}{2k+1}
\left\langle\overline{R_{3d}(r)}\overline{j_0}\left(\pi \frac{r'}{R}\right)\left|
\frac{r^k}{r'^{k+1}}\right|j_0\left(\pi \frac{r}{R}\right)R_{3d}(r')\right\rangle 
\nonumber\\ \mbox{} \nonumber \\ 
& & \times \left[ \left\langle 
\overline{Y_2^1}\overline{Y_0^0}\left|\sum_{m=-k}^kY_k^m(1)\overline{Y_k^m}(2)\right|
Y_0^0Y_2^{1} \right\rangle \right.  \nonumber \\
\mbox{} \nonumber \\
& & + \left\langle 
\overline{Y_2^1}\overline{Y_0^0}\left|\sum_{m=-k}^kY_k^m(1)\overline{Y_k^m}(2)\right|
Y_0^0Y_2^{-1} \right\rangle  \nonumber \\
\mbox{} \nonumber \\
& & + \left\langle 
\overline{Y_2^{-1}}\overline{Y_0^0}\left|\sum_{m=-k}^kY_k^m(1)\overline{Y_k^m}(2)\right|
Y_0^0Y_2^{1} \right\rangle  \nonumber \\
\mbox{} \nonumber\\
& & + \left. \left\langle 
\overline{Y_2^{-1}}\overline{Y_0^0}\left|\sum_{m=-k}^kY_k^m(1)\overline{Y_k^m}(2)\right|
Y_0^0Y_2^{-1} \right\rangle \right]\sum_{a_i, a'_i}{w_{1S}(r-a_i)w_{1S}(r'-a'_i)}
\end{eqnarray}

\noindent
The conditions  for non-zero Clebsch-Gordan coefficients give the selection rules for
k and m, and also 
determine the terms of the sum in the radial part. 
For the exchange interaction k can be only 2. And the radial part of the matrix element becomes:
\begin{equation}
\rho^{Exchange}(3d,s,k=2) = \int r^2dr\int r'^2dr' R_{3d}(r)
R_{1s}^{Dot}(r')\frac{r^2}{r'^3} R_{1s}^{Dot}(r)R_{3d}(r')\sum_{a_i, a'_i}{w_{1S}(r-a_i)w_{1S}(r'-a'_i)}
\end{equation}

\noindent
Here $R_{1s}^{Dot}(r)$ is the envelope function for the electron
confined in a dot. 
$R_{3d}$ is the orbital function of the 3d-electron, which can be
approximately taken as the radial Slater function\cite{GR}
$\normalsize\ R_{3d}(r) = r^2e^{-1.87r/a}$, where $ a=\frac{\hbar^2}{me^2}$
The sum over the lattice sites will be normalized so as the result we have 
\begin{equation}
\rho^{Exchange}(3d,s,k=2) = \int r^2dr\int r'^2dr'R_{3d}(r)R_{1s}^{Dot}(r')
\frac{r^2}{r'^3}R_{1s}^{Dot}(r)R_{3d}(r')
\end{equation}

\noindent
and for the Coulomb interaction the only possible value of k 
is $0$ and the radial part
of the matrix element has the form:
\begin{equation}
\rho^{Coulomb}(3d,s,k=2) = \int r^2dr\int r'^2dr' R_{3d}(r)
R_{1s}^{Dot}(r')\frac{1}{r'} R_{3d}(r)R_{1s}^{Dot}(r')
\end{equation}

\noindent
The integrals  are taken over the dot whose radius is R.
Then the radial integral $\rho(3d,s,k)$ (28) can be computed:
\begin{eqnarray}
\rho^{Coulomb}(3d,s,k=0) & = & \frac{3.96}{R} \nonumber \\
\rho^{Exchange}(3d,s,k=2) & = & \frac{240[-3(\frac{1.87}{a})^5\frac{\pi}{R}+10(\frac{1.87}{a})^3(\frac{\pi}{R})^3
-3\frac{1.87}{a}(\frac{\pi}{R})^5]}{[(\frac{1.87}{a})^2 +(\frac{\pi}{R})^2]}
\end{eqnarray}

\noindent
As the result of the calculation we have for the matrix elements of the Coulomb and Exchange interactions between 
the 
extra electron 
and the $^6A_1$ state:
\begin{eqnarray}
K(^6A_1,^1S) & = & 5e^2\rho^{Coulomb}_{k=0} = K, \nonumber\\
J(^6A_1,^1S) & = & \frac{e^2}{25}\rho^{Exchange}_{k=2} = J
\end{eqnarray}

\noindent
And for the Coulomb and Exchange interaction between the extra electron and the $^4T_1$ states:
\begin{eqnarray}
K(^4T_1,^1S) & = & 5e^2\rho^{Coulomb}_{k=0} = K, \nonumber\\
J(^4T_1xy,^1S) & = & \frac{e^2}{25}\rho^{Exchange}_{k=2} = J \nonumber \\
J(^4T_1z,^1S) & = & -\frac{e^2}{25}\rho^{Exchange}_{k=2} = -J \nonumber \\
\end{eqnarray}

\noindent
And as the effect of the Coulomb and exchange interaction  with the injected electron,
the levels $^6A_1$  will be shifted and split into 2 sublevels with 
total spin $S=3$ and $S=2$ with the corresponding exchange energy:
\begin{eqnarray}
E_{ex}(^6A_1,\varphi(1s,+), S=3) & = & K -\frac{7}{4} J \nonumber \\
\mbox{} \nonumber\\
E_{ex}(^6A_1,\varphi(1s,-), S=2) & = & K + \frac{5}{4} J
\end{eqnarray}

\noindent
with the corresponding wavefunctions:
\begin{eqnarray}
|^6A_1,\varphi(1s,+), S=3> & = & [-\xi^+\eta^+\zeta^+u^+v^+]|s+> \nonumber \\
|^6A_1,\varphi(1s,+), S=2> & = & [-\xi^+\eta^+\zeta^+u^+v^+]|s->
\end{eqnarray}

\noindent
And the $^4T_1$ levels splits into four sublevels: two sublevels with the 
value of the total spin $S=2$:
\begin{eqnarray}
E_{ex}(^4T_1,z\varphi(1s,+), S=2) & = & K + \frac{3}{4} J \nonumber \\
\mbox{} \nonumber\\
E_{ex}(^4T_1,xy\varphi(1s,+), S=2) & = & K - \frac{3}{4}J \nonumber \\
\end{eqnarray}

\noindent
and the two sublevels with the total spin $S=1$
\begin{eqnarray}
E_{ex}(^4T_1,z\varphi(1s,-), S=1) & = & K - \frac{5}{4}J \nonumber \\
\mbox{} \nonumber \\
E_{ex}(^4T_1,xy\varphi(1s,+), S=2) & = & K + \frac{5}{4} J \nonumber \\ 
\end{eqnarray}

\noindent
with the corresponding wavefunctions:
\begin{eqnarray}
|^4T_1z,\varphi(1s,+), S=2> & = & [\xi^+\eta^+\zeta^+\zeta^-u^+]|s+> 
+ \frac{\beta}{\sqrt{2}}[\xi^2\eta^+\zeta^+p_y^+]|s+> 
-\frac{\beta}{\sqrt{2}}[\xi^+\eta^2\zeta^+p_x^+]|s+> \nonumber
\mbox{} \nonumber \\
|^4T_1xy,\varphi(1s,+), S=2> & = & 
\frac{\sqrt{2}}{4}[\xi^+\xi^-\eta^+\zeta^+u^+]|1s+)  
+ \frac{\sqrt{6}}{4}[\xi^+\xi^-\eta^+\zeta^+v^+]|1s+> \nonumber 
\mbox{} \nonumber \\
&&- \frac{1}{2}[\xi^+\eta^+\eta^-\zeta^+u^+]|1s+> 
+ [\xi^+\eta^+\eta^-\zeta^+v^+])|1s+> 
+ \frac{\beta}{2}[\xi^+\eta^2\zeta^+p_z^+]|1s+> \nonumber
\mbox{} \nonumber \\
&&- \frac{\beta}{2}[\xi^+\eta^+\zeta^2p_y^+]|1s+>
+\frac{\beta}{2}[\xi^2\eta^+\zeta^+p_z^+]|1s+> -
\frac{\beta}{2}[\xi^+\eta^+\zeta^2p_x^+]|1s+> \nonumber \\
\end{eqnarray}

\noindent 
and
\begin{eqnarray}
|^4T_1z,\varphi(1s,-), S=1> & = & [\xi^+\eta^+\zeta^+\zeta^-u^+]|s->
+ \frac{\beta}{\sqrt{2}}[\xi^2\eta^+\zeta^+p_y^+]|s-> -
\frac{\beta}{\sqrt{2}}[\xi^+\eta^2\zeta^+p_x^+]|s-> \nonumber
\mbox{} \nonumber \\
|^4T_1xy,\varphi(1s,-), S=1> & = &
\frac{\sqrt{2}}{4}[\xi^+\xi^-\eta^+\zeta^+u^+]|1s-)
+ \frac{\sqrt{6}}{4}[\xi^+\xi^-\eta^+\zeta^+v^+]|1s-> \nonumber 
\mbox{} \nonumber \\
&&- \frac{1}{2}[\xi^+\eta^+\eta^-\zeta^+u^+]|1s-> 
+ [\xi^+\eta^+\eta^-\zeta^+v^+])|1s-> 
+ \frac{\beta}{2}[\xi^+\eta^2\zeta^+p_z^+]|1s-> \nonumber
\mbox{} \nonumber \\
&&-\frac{\beta}{2}[\xi^+\eta^+\zeta^2p_y^+]|1s-> 
+\frac{\beta}{2}[\xi^2\eta^+\zeta^+p_z^+]|1s-> 
-\frac{\beta}{2}[\xi^+\eta^+\zeta^2p_x^+]|1s-> \nonumber \\
\end{eqnarray}

\noindent
The two sublevels $|^4T_1z,\varphi(1s+), S=2>$ and $|^4T_1 xy,\varphi(1s+), S=2>$ of
the $^4T_1$ state of the $Mn^{2+}$ ion and the sublevel $|^6A_1,\varphi(1s-), S=2>$
now have the same total spin $S=2$ and the transitions between them now 
are allowable.
In Fig.2 we show the exchange energy splitting of the $^4T_1$ and $^6A_1$ 
levels and the transitions allowable between them.

So the exchange interaction between the electron injected in the dot and the d- 
electrons of the Mn-impurity center really make the previously forbidden transition 
become allowable.
In the next section we will give some numerical calculations for the transition 
probability and the transition lifetime.
\vspace{0.5cm}

{\bf VI. Luminescence and the Lifetime of the Transition}

In this section we will give some numerical calculations for the transition
probability and the transition lifetime.

The electric dipole transition matrix element between two states $|i> $ 
and $|j>$ is written in the form\cite{SO}:
\begin{equation}
<i|r.\hat{\epsilon}|j> = 
\frac{4}{3}\sum_q\left\langle k \left| 
r.Y^q_1\left(\frac{\vec{r}}{r}\right)\right|s\right\rangle \times Y_1^{-q} 
\left(\frac{\vec{\epsilon}}{\epsilon}\right) \end{equation}
\noindent
So the matrix element of the transition between the sublevels of the state $^4T_1$ and
$^6A_1$ will again have the form of the determinantal matrix element between the 
wavefunctions (40) and (43).
Again, here we can separate the radial and the angular part 
\begin{eqnarray}
& & \left\langle^4T_1z z,\varphi^{Dot}_{1s+}, S=2 \left| r.\hat{\epsilon} \right| 
^6A_1,\varphi^{Dot}_{1s-}\right\rangle
\nonumber \\
\mbox{} \nonumber\\
& &= \left\langle R_{3d, 4p}R_{1s}^{Dot}\left| r 
\right|R_{3d,4p}R_{1s}^{Dot}\right\rangle \left\langle(i)\left. Y^q_1 \right |(j)\right 
\rangle 
\end{eqnarray}

\noindent
where $<(i)|r|(j)>$ denote the angular part and the $< |r|>-$ the radial part.
Similar to 
the calculation of the exchange matrix element in the last section, the angular part
is calculated using the Clebsch-Gordan coefficients. The radial part also can be computed
using the expansion of Bessel function of the dot envelope function $R_{1S}^{Dot}$.

Because the interaction with the extra electron releves the spin forbideness, the transition between the 
$^4T_1$ and $^6A_1$ states with the same spin $S=2$ will be allowed. Note again that the electric dipole 
transition is 
allowed due to mixing of the 4p-state into the 3d-states, and depend on the small coefficients $\beta$.
The matrix element (46) will be based on the one-electron matrix element $<3d|r|4p>$, which is allowed
with the angular part $<Y^l_2|Y^q_1|Y^{l'}_2>$. The radial part show a comlicated dependance on the inverse 
of the dot radius and its powers.
 
We do some numerical calculations for the case of a quantum dot of Mn- doping 
ZnS . We use here the standard parameters of Mn and ZnS.
For the doped ZnS:Mn quantum dot of radius
50\AA, we obtain the
splitting due to exchange will be 1.0 to 1.5 eV, which is rather large.
The small coefficient $\beta$ of the parity mixing  is reported about $2.6-3\time 10^{-2}$, 
then the probability 
of the transition approximately equals $10^{-3}$( the oscillator strength of allowed 
electric dipole transition has magnitude of 1).
It results in a lifetime, that is inversely proportional to the transition probability. The 
transition lifetime  in our calculation is approximately $2\times 10^{-5} s$.
The integral strongly inversely depends on the radius of the nanocrystal.
For the dot of the radius of 40\AA, the transition lifetime is $1.6\times 10^{-5} s$, and
it is $0.9\times 10^{-5} s$ for the dot of radius 30\AA.
For a smaller dots, the matrix element will be larger, the probability of the transition 
will be higher and the transition life-time is shorter.

To compare with the experiments, we notice that our result for the transition lifetime
for this size of dot is two orders larger
than the transition lifetime in the Mn doped ZnS bulk (1.8 ms). So the  
presence of an extra electron in the quantum dot really makes the spin-forbidden transition  
allowable and shortens the life time. The theory supports the results obtained in experimental
works [1-4, 8-12] and is close to the result in [10] although we do not
obtain the 5 orders of magnitude lifetime shortening, which is reported by
R.N. Bhargava et al.[1-4].

{\bf Summary}

   In this work, we presented a theory for the new model to control the 
optical transitions in the nanocrystal. By injecting one extra electron into the dot, one 
can change the transition probability and the optical properties of the nanocrystal. 
The exchange interaction between the extra electron on the impurity 
electron splits the energy levels and makes the former spin-forbidden transition become allowable 
and decreases the lifetime of the transition by about two orders of 
magnitude. 

We are grateful to Dr. Al. L. Efros for his helpful suggestion and 
discussions and Dr. D. Norris for helpful conversation. We also 
acknowledge partial support from NYSTAR ECAT Project $N^o0000067$ and 
PSC-CUNY for this work.

\newpage
{\bf FIGURE CAPTIONS}
\vspace{2cm}

\noindent
{\bf Fig.1}. Splitting diagram of a single 3d electron in an octahedral crystal field. 
The five-fold degenerate d-electron with 
energy $E_0$ splits in an octahedral crystal field into two levels. \\
\vspace{0.7cm}

\noindent
{\bf Fig.2}.Exchange energy splitting of the $^4T_1$ and $^6A_1$ states of
the $Mn^{2+}$ ion. Arrows indicate the allowable transitions.\\

\end{document}